\documentclass{emulateapj}

\bibpunct{(}{)}{;}{a}{}{,}

\begin{document}
\title{Disentangling Visibility and Self-Promotion Bias in the
  arXiv:astro-ph Positional Citation Effect} 

\author{J.\,P. Dietrich} 
\affil{ESO, Karl-Schwarzschild-Stra{\ss}e 2, 85748 Garching b. M\"unchen,
  Germany} 
\email{jdietric@eso.org}

\begin{abstract}
  We established in an earlier study that articles listed at or near
  the top of the daily arXiv:astro-ph mailings receive on average
  significantly more citations than articles further down the list. In
  our earlier work we were not able to decide whether this positional
  citation effect was due to author self-promotion of intrinsically
  more citable papers or whether papers are cited more often simply
  because they are at the top of the astro-ph listing.
  Using new data we can now disentangle both effects.
  Based on their submission times we separate articles into a
  self-promoted sample and a sample of articles that achieved a high
  rank on astro-ph by chance and compare their citation distributions
  with those of articles in lower astro-ph positions.
  We find that the positional citation effect is a superposition of
  self-promotion and visibility bias.
\end{abstract}

\section{Introduction}
\label{sec:introduction}
In \citet[][Paper
I]{2008PASP..120..224D}\defcitealias{2008PASP..120..224D}{Paper I} we
studied the effect of an e-Print's placement in the daily
arXiv:astro-ph listing on the number of citation it gets. We found
that e-Prints appearing at or near the top of the astro-ph mailings
receive significantly more citation than articles further down the
list. We proposed three non-exclusive effects to explain this
\emph{positional citation effect} (PCE). These are defined as:
\begin{itemize}
\item The Visibility Bias (VB) postulate -- Papers appearing at the
  top of the astro-ph listing are seen by more people and thus cited
  more often than those further down the list, where the attention of
  the astro-ph readers might decrease;
\item The Self-promotion Bias (SP) postulate -- Authors tend to
  promote their most important works, and thus most citable articles,
  by placing them at prominent positions;
\item The Geography Bias (GB) postulate -- The submission deadline
  preferentially puts those authors at the top of the listing whose
  working hours coincide with the submission deadline. This group
  already has higher citation counts for other reasons.
\end{itemize}
The last postulate pertains to the facts that (1) US American authors have
a higher fraction of highly cited papers than their European
colleagues \citep{2007EurRev..15..3S} and (2) the arXiv submission
deadline of 16:00~EST/EDT is within the normal working hours of
astronomers in the US, while it is not for European astronomers.

We concluded in \citetalias{2008PASP..120..224D} that GB is not the
cause of the observed PCE because the effect is found independently in
the samples of European and US authors. We proposed to disentangle VB
and SP by the following method: using the submission times of e-Prints
and grouping them into two samples, one that is submitted so shortly
after the deadline that it is statistically expected to be
self-promoted, and a second one that is submitted long enough after
the deadline to exclude self-promotion, and repeating the citation
analysis for both samples, one can distinguish between SP and VB.
According to information we received at the time of writing
\citetalias{2008PASP..120..224D} from arXiv administrators, the
initial submission times are not stored. Consequently, we were not
able to decide whether the PCE is caused by VB or SP. Meanwhile we
were contacted by arXiv staff informing us that, in fact, the original
submission times, although indeed not stored as part of an e-Prints
record, can be recovered from the server log files. This now enables
us to perform the timing analysis proposed in
\citetalias{2008PASP..120..224D}.

\section{Analysis}
\label{sec:analysis}
We use the same sample as in \citetalias{2008PASP..120..224D}, i.e.,
astro-ph e-Prints published between the beginning of July 2002 and the
end of December 2005. Citation data for these e-Prints were gathered
from NASA's ADS Bibliographic Services\footnote{Access this service
  through http://adsabs.harvard.edu/index.html}. We do not correct for
the fact that older papers had more time to gather citations than
e-Prints published towards the end of the period under investigation
here. For every astro-ph e-Print published in one of the core journals
of Astronomy (in agreement with \citet{2005IPM....41.1395K} we define
these as \emph{The Astrophysical Journal (Letters)} and its
\emph{Supplement Series}, \emph{Astronomy \& Astrophysics},
\emph{Monthly Notices of the Royal Astronomical Society}, \emph{The
  Astronomical Journal}, and \emph{Publications of the Astronomical
  Society of the Pacific}) we compute the time $t_\textrm{s}$ passed
from the last submission deadline to the submission time of the
article to the arXiv server. We ruled out GB as the sole cause of the
PCE in \citetalias{2008PASP..120..224D} but we now restrict our
analysis to articles whose first author's first affiliation is in
North or South America. The reasons for this choice are that European
authors must self-promote to achieve the top position on astro-ph,
weakening any VB signal if present, and to avoid any residual signal
from GB.  Restricting our analysis to American (North and South)
authors we avoid a correlation of different citation distributions
with submission behavior while at the same time maximizing the sample
size.

\begin{figure}
  \plotone{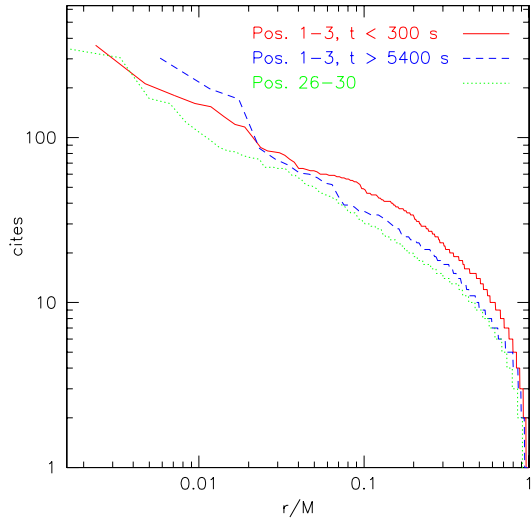}
  \caption{Zipf plots for the timing analysis. The $x$-axis show the
    normalized rank of astro-ph postings in their respective samples
    after sorting them by citations. The $y$-axis gives the number of
    citations. The different color/line-styles represent the different
    samples under investigation. The solid red line represents
    e-Prints in the first three astro-ph positions posted within the
    first 5 minutes after starting a new list. The dashed blue line is
    the Zipf law for articles in the same positions but posted more
    than 1.5\,h after the deadline. The dotted green line gives all
    articles of American authors in positions 26--30 for comparison.}
  \label{fig:zipf-timing}
\end{figure}

We perform an analysis of the citation counts similar to the one in
\citetalias{2008PASP..120..224D}. The citation distribution is a power
law \citep{1998EPJB....4..131R}, which is best analyzed using a Zipf
plot. A Zipf plot shows the citations on the $r^\mathrm{th}$ most
cited paper out of an ensemble of size $M$ versus its rank $r$ or, if
several samples of different sizes are to be compared, its normalized
rank $r/M$.  Figure~\ref{fig:zipf-timing} shows Zipf plots for three
different samples of core journal articles; two samples of articles in
the first three astro-ph positions, one submitted very shortly after
the deadline ($t_\mathrm{s}<300$\,s) very likely aimed at
self-promotion and one submitted obviously without the intent to
achieve the ``pole position'' ($t_\mathrm{s} > 5400$\,s). We bin the
first three positions because the PCE is much for stronger for them
than for lower positions at which it is still significant and to
average out the noise that would dominate in studying individual
positions.  The third sample contains articles that appeared at
astro-ph positions 26--30 at any submission time.

We clearly see that the three curves, while their slopes are roughly
equal, are at different loci, corresponding to different
normalizations of the citation distribution power law. The highest
curve, i.e. the highest normalization of the citation power law is the
sample of articles submitted shortly after the deadline. Articles
listed at the top positions but submitted later are cited less often,
with the exception of the three most cited articles in this sample,
but still considerably more often than articles further down the list.

To quantify the impact of VB and SP we compute the average citations a
paper gets in the range $-3.0 < \ln(r/M) < -1.0$. We choose this range
to avoid the bulk of mostly ignored papers and especially to avoid
being dominated by a few highly cited papers. We find that articles in
the early sample are on average cited $34.4\pm1.1$ times, while
articles from the later sample are cited $26.2^{+1.3}_{-1.4}$ times.
The comparison sample from astro-ph positions 26--28 has a mean
citation number of $22.0\pm0.7$. The quoted errors are the 68\%
confidence intervals estimated from bootstrap resampling the citation
counts in the selected interval. 

\begin{figure}
  \plotone{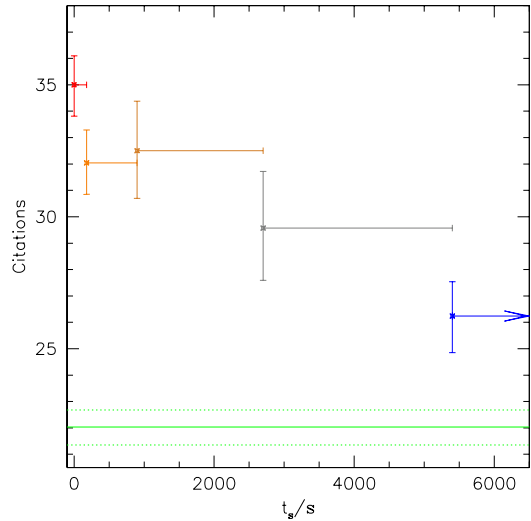}
  \caption{Average citation numbers for articles in the first three
    positions of astro-ph depending on their submission time. The
    horizontal error bars denote the width of the sample bins. The
    green horizontal line corresponds to the average citations (with
    dotted error bars) of e-Prints at positions 26--30.}
  \label{fig:timing}
\end{figure}

We repeat this calculation for three additional time intervals and
plot the result in Fig.~\ref{fig:timing}. We find that after the
initial rush of self-promoted papers the citation rates slowly drop
for e-Prints submitted later after the deadline. This confirms a
contribution of SP to the PCE. We also find that papers submitted more
than 1.5\,h after the deadline, i.e., those e-Prints that achieved a
high position in the astro-ph listing almost certainly by chance and
not by the submitter's intent, are still cited significantly more
often ($3\sigma$) than papers further down the list. This proves that
also VB contributes to the PCE. We note that these results are
independent of the exact binning that is employed. Choosing different
bins close to and far away from the submission deadline moves the
points in Fig.~\ref{fig:timing} somewhat up and down. The overall
result that papers submitted shortly after the deadline have higher
citation rates than e-Prints submitter later remains, as does the
difference in citation numbers between late articles at the top and
articles down the astro-ph listing. The presence of both SP and VB
thus does not depend on the exact binning employed.

\section{Summary and Discussion}
\label{sec:concl}
We studied the factors contributing to the increased number of
citations e-Prints at the top of the daily astro-ph listing receive
compared to e-Prints listed further down the astro-ph mailing. By
making a timing argument we constructed samples of e-Prints appearing
at the three first positions of astro-ph that are either (1) almost
certainly submitted with the intent of getting the top spot; or (2)
achieved a high position of astro-ph purely by chance; or (3) fall
somewhere in between and have a mixture of categories.

We found that the sample of self-promoted papers indeed has the
highest citation rates. This shows that self-promotion as a mechanism
that preferentially puts intrinsically more citable papers to the top
of the astro-ph listings in fact works. This is not surprising,
considering that \citet{2005IPM....41.1395K} found evidence for a
self-selection bias in which papers authors post on astro-ph. We, in
turn, find a similar effect within the e-Prints on astro-ph.

Arguably, the more important finding of this work is the difference in
citation rates between the not self-promoted sample of e-Prints and
articles appearing much lower in the astro-ph mailing. The citation
rates for the late sample are lower than for the self-promoted sample
but still significantly higher than for articles lower in the astro-ph
mailing. This provides strong evidence for the visibility bias theory
that articles are cited more often, not due to some inherent quality
they have, but simply because they are at the top of the astro-ph
listing.

Citation counts are often used to evaluate the scientific quality of
individuals or institutions and hiring or funding decisions are partly
based on them. Our finding that a visibility bias exists at the top of
arXiv:astro-ph should provide a strong cautionary note concerning the
use of such statistics. We also note that the fraction of astro-ph
e-Prints submitted very shortly after the deadline increased during
the interval under study here. In the second half of 2002 1.5\%
(2.9\%) of all e-Prints were submitted within the first 60\,s
(300\,s). In the second half of 2005 these numbers rose to 2.3\%
(4.6\%). The chance that this is a statistical fluctuation is smaller
than 0.01\%. This change in submission behavior appears to be
indicative of a growing feeling in the astronomical community that VB
plays a role and that citations are not awarded purely on merit of the
work presented in a paper.

One could simply get rid of VB by randomizing the order of the
astro-ph listing for every reader. In this way the VB would average
out over the readership of astro-ph. However, by doing so one would
ignore the underlying problem that leads to VB in the first
place. Everyday astronomers are confronted with an enormous amount of
new information, which they have to sort, classify, and ultimately
decide what is of relevance for their own research. 

Publications are never cited without a reason, i.e., VB does not lead
to unjustified additional citations of a paper. Thus, we must draw the
conclusion that papers down the astro-ph list are overlooked and not
cited when they should be. Any randomization would mitigate the VB
problem by changing the set of papers that does not get the attention
it deserves but it would not fix the real problem, i.e., the
information overload which astronomers face every day. It is important
to realize that the VB effect on citations is only a secondary
effect. The primary effect, from which the citation inequality
follows, is that researchers are not aware of relevant publications
and results in their own field. This is potentially an impediment
for science and the real problem that needs fixing. Since we cannot
expect the number of publications to decrease, the solution has to be
in the way information is presented. 

Only a relatively small subset of e-Prints is relevant to any
individual researcher and the daily challenge is to identify these in
the much larger astro-ph listing. A possible first step in this
direction is the arxivsorter \citep{2007Arxivsorter..M}, which we
already mentioned in \citetalias{2008PASP..120..224D}. Arxivsorter
aims to sort daily, recent, or monthly astro-ph listings by relevance
to an individual reader. The underlying idea is that scientists
through co-authorship form an interconnected network of authors. By
specifying a few authors relevant to a reader's fields of interest,
the ``proximity'' of a new e-print in the author network can be
calculated. This proximity seems to be a good proxy for relevance to a
reader's interests.

\acknowledgements The original submission times of e-Prints were
provided by Paul Ginsparg. I thank Bruno Leibundgut, Brice M\'enard,
Uta Grothkopf, and the anonymous referee for comments that helped to
improve the manuscript. This research has made use of NASA's
Astrophysics Data System Bibliographic Services.

\end{document}